\documentclass[runningheads]{llncs}
\DeclareUnicodeCharacter{FF1A}{:}
\usepackage[T1]{fontenc}
\usepackage{graphicx}
\usepackage{multirow}
\usepackage{array}
\usepackage{algorithm}
\usepackage{algorithmic}
\usepackage{booktabs}
\usepackage{amssymb}
\begin{document}
\title{SAM-Guided Robust Representation Learning for \\One-Shot 3D Medical Image Segmentation}
%
\author{Jia Wang \and
Yunan Mei \and
Jiarui Liu\and Xin Fan\thanks{Corresponding author}}
\institute{School of Software Technology, Dalian University of Technology, Dalian, China
}
\maketitle              
%
\begin{abstract}
One-shot medical image segmentation (MIS) is crucial for medical analysis due to the burden of medical experts on manual annotation. The recent emergence of the segment anything model (SAM) has demonstrated remarkable adaptation in MIS but cannot be directly applied to one-shot medical image segmentation (MIS) due to its reliance on labor-intensive user interactions and the high computational cost. To cope with these limitations, we propose a novel SAM-guided robust representation learning framework, named RRL-MedSAM, to adapt SAM to one-shot 3D MIS, which exploits the strong generalization capabilities of the SAM encoder to learn better feature representation. We devise a dual-stage knowledge distillation (DSKD) strategy to distill general knowledge between natural and medical images from the foundation model to train a lightweight encoder, and then adopt a mutual exponential moving average (mutual-EMA) to update the weights of the general lightweight encoder and medical-specific encoder. Specifically, pseudo labels from the registration network are used to perform mutual supervision for such two encoders. Moreover, we introduce an auto-prompting (AP) segmentation decoder which adopts the mask generated from the general lightweight model as a prompt to assist the medical-specific model in boosting the final segmentation performance. Extensive experiments conducted on three public datasets, i.e., OASIS, CT-lung demonstrate that the proposed RRL-MedSAM outperforms state-of-the-art one-shot MIS methods for both segmentation and registration tasks. Especially, our lightweight encoder uses only 3\% of the parameters compared to the encoder of SAM-Base.
\keywords{SAM  \and One-shot Segmentation \and Representation Learning.}
\end{abstract}
\section{Introduction}

\begin{figure*}[!t]
\centering
\setlength{\belowcaptionskip}{-3mm}
\includegraphics[scale=0.15]{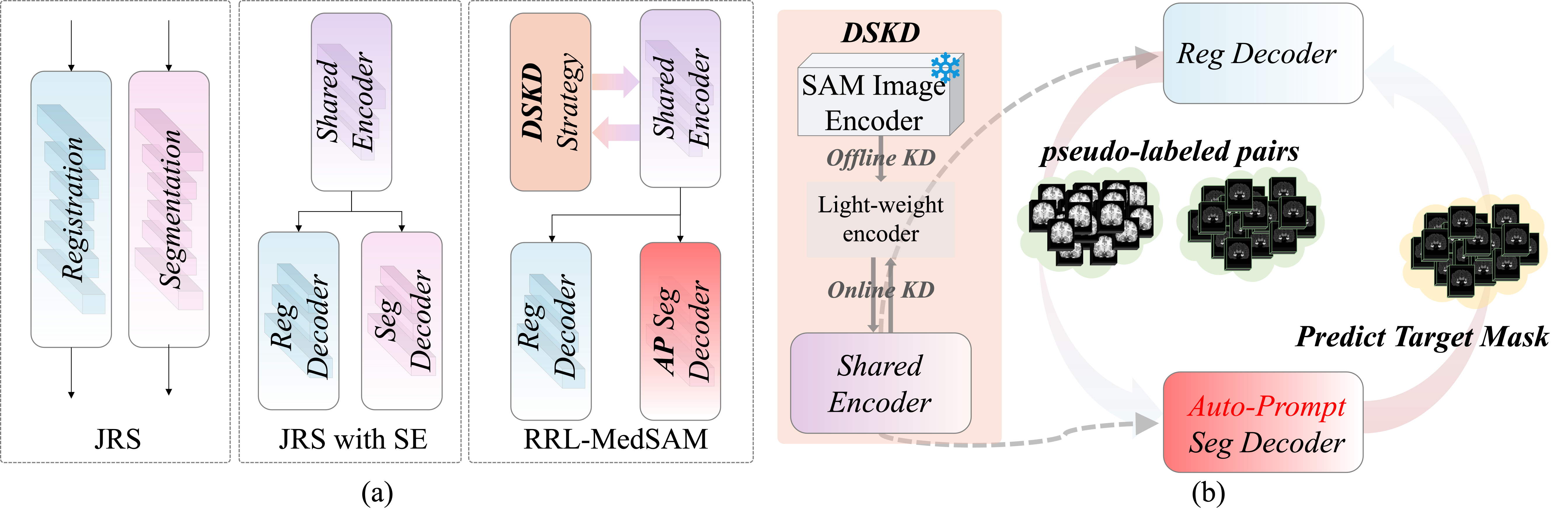}
\vspace{-0.9cm}
\caption{(a) Framework comparison between our method and previous one-shot MIS methods. (b) A simplified schematic of the RRL-MedSAM framework.  ``KD'' denotes knowledge distillation. ``SE" denotes a shared encoder.}
\label{fig1}
\end{figure*}

Accurate medical image segmentation (MIS) is crucial for various medical applications, including tumor detection, radiotherapy planning, and cardiac imaging \cite{elmahdy2021joint,sinclair2022atlas,vandewinckele2020segmentation}. 
One-shot MIS \cite{zhao2023one,ding2021modeling,he2022learning,sun2023dual}  has emerged as a promising approach to reduce the heavy reliance on extensively annotated datasets, which uses a single annotated sample reducing the need for extensive labelling. One paradigm is Joint Registration and Segmentation (JRS) utilizing the voxel-wise correspondence between an atlas image and an unlabeled image through registration (see
the left part of Fig. 1 (b)), allowing the warped atlas label to serve as a pseudo label for the target image. However, these methods suffer from structural redundancy due to the use of independent encoder-decoder frameworks. Shared encoder and task-specific decoder structures \cite{fan2024bi,andresen2022deep,zhao2021deep} (see
the middle part of Fig. 1 (b)) have been shown to improve speed and accuracy, but the limited feature representation of the shared encoder can hinder segmentation performance.

The Segment Anything Model (SAM) \cite{kirillov2023segment}, a powerful tool in weakly-supervised image segmentation, has demonstrated strong robustness in segmenting natural images. Recent efforts to adapt SAM to medical imaging have focused on parameter-efficient methods \cite{gong20233dsam,wu2023medical}, validating its potential for medical segmentation. However, SAM still faces three significant challenges when applied to one-shot MIS: (i) the reliance on labor-intensive user inputs, (ii) discrepancies in data distribution between natural and medical images, and (iii) the high computational demands of the SAM encoder, which limit its real-world clinical applicability.

To address these challenges, we propose RRL-MedSAM, a novel SAM-guided robust representation learning framework (see
the right part of Fig. 1 (b)). Our method adapts SAM to one-shot 3D medical segmentation by utilizing the SAM encoder’s generalization capabilities. The framework consists of two parts: a lightweight model that distills common image knowledge from a large encoder, and a medical-specific model that uses JRS with a shared encoder to perform one-shot medical segmentation. We introduce a dual-stage knowledge distillation strategy to transfer knowledge between natural and medical images and use mutual exponential moving average (mutual-EMA) to update the encoders’ weights. Pseudo labels from the registration network guide mutual supervision between the two encoders, and the mask generated from the lightweight model assist the medical-specific model, improving segmentation accuracy.

\begin{figure*}[!t]
\centering
\includegraphics[scale=0.34]{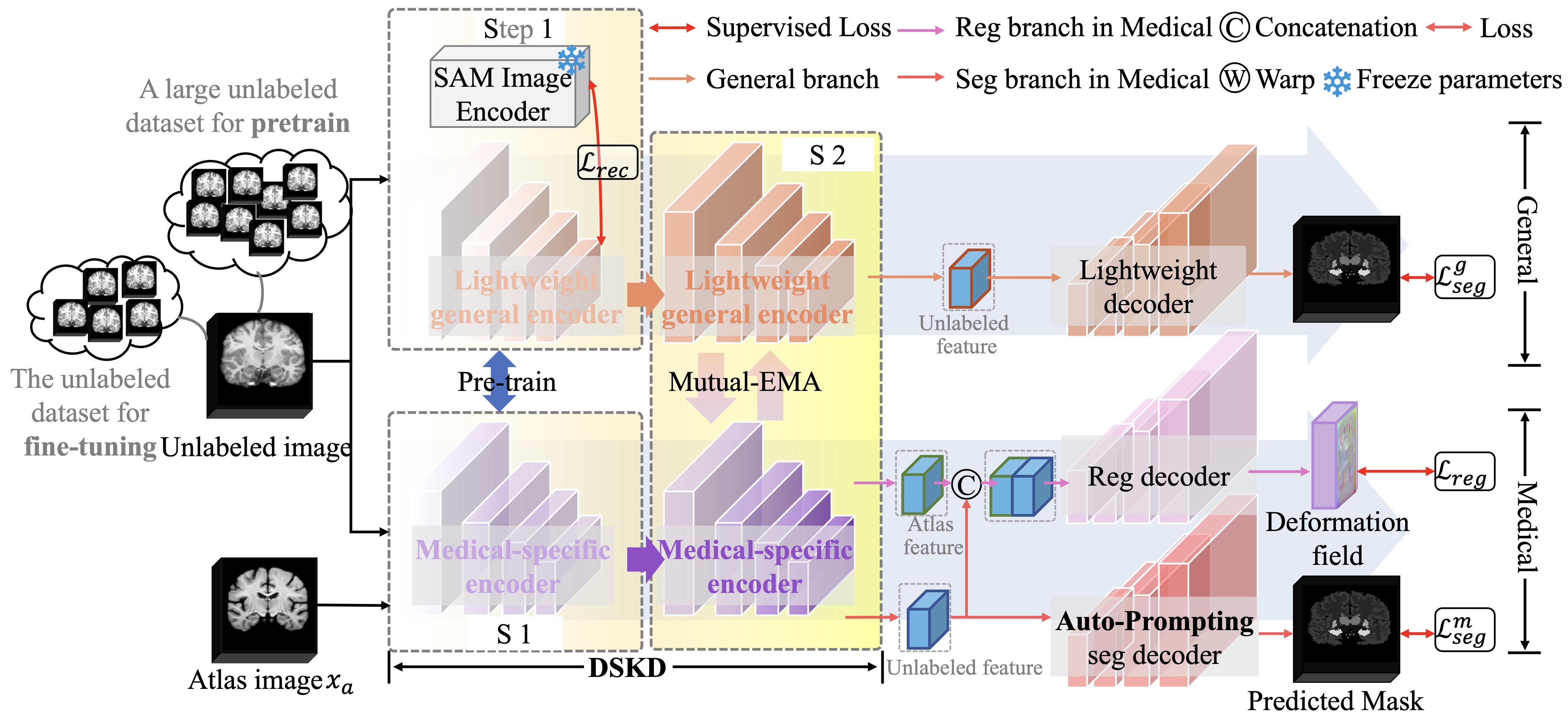}
\vspace{-0.4cm}
\caption{Overview of the proposed RRL-MedSAM. The network consists of general and medical branches, which learn the common image and medical-specific knowledge, respectively. }
\label{framework}
\vspace{-0.6em}
\end{figure*}

\section{Methodology}
\subsection{Overview of the Proposed Framework}
We propose a SAM-guided robust representation learning framework, named RRL-MedSAM, to adapt SAM to one-shot 3D medical image segmentation. Our main idea is to exploit the strong generalization capabilities of the SAM image encoder to learn better feature representation. Denote an atlas image and its segmentation labels as ($x^a, y^a$), a large number of unlabeled images as $D_{pre} = \{x_i^U\}_{i=1}^{N}$ for the pre-train general lightweight encoder and the medical-specific encoder, and $D_{fin} = \{x_i^u\}_{i=1}^{M}$ for subsequently fine-tune stage.

The overall framework is shown in Fig. \ref{framework}, consisting of two branches: a general encoder-decoder branch for learning general knowledge and a medical branch with a shared encoder and task-specific decoder for medical knowledge. The training process includes two phases: we first train the lightweight general encoder by SAM-guided and the medical-specific encoder with datasets $D_{pre}$. Then we employ the pre-trained encoder to fine-tune the final model with datasets $D_{fin}$ detailed in the next section. Moreover, the pre-trained encoder can serve as a plug-and-play feature extractor for fine-tuning any site brain datasets. Notably, the same architecture and encoder parameters are shared across both branches during pre-training and fine-tuning.

\subsection{Dual-Stage Knowledge Distillation}
The large number of parameters and the high computational costs of SAM limit its applications in real-world clinical settings. Recent methods \cite{xiong2024efficientsam,zhang2024efficientvit,xu2024esp} aim to achieve model compression by transferring the knowledge of the SAM encoder to a lightweight encoder learned from its predictions. However, this one-stage strategy struggles with harmonizing feature representations across diverse modalities, particularly for 3D medical images. \textbf{Due to the particularity of our method, we not only need to tackle the domain gap but also ensure harmonization of the feature representation between registration and segmentation.} This means that we need an encoder with powerful feature representation capability.

To achieve this, we propose a dual-stage knowledge distillation (DSKD) strategy for robust representation learning. It is well-known that there exists general knowledge between natural images and medical images as shown in Fig. \ref{fig1} (c). 

\textbf{The first stage} transfers common image knowledge of the foundation teacher model into a lightweight general encoder. To be more specific, we adopt the mean squared error (MSE) similarity to compare the outputs from the SAM image encoder and the lightweight general encoder. Let us denote the SAM image encoder as $f_{sam}$, and the lightweight general encoder as $f_{lie}$. Moreover, the target features from $f_{sam}$ can be written as $f_{sam}(x) = f_{sam}(\{x_i\}_{i=1}^N)$. In the same way, the output from a lightweight general encoder is $f_{lie}(x)$. The above process can be summarized as 
\begin{equation}
    \mathcal{L}_{rec} = \frac{1} {N} \cdot\sum_{i=1}^N \Vert f_{sam} - f_{lie} \Vert,
\end{equation}
where $N$ is the number of input batches, $\Vert \cdot \Vert$ denotes a norm. In addition, the medical-specific encoder is trained simultaneously by the JRS with a shared encoder framework. 

With pre-trained lightweight general and medical-specific encoders, \textbf{the second stage} introduces a mutual-EMA strategy for parameter-level interaction and shared knowledge. In addition to the EMA from $f_{sam}$ to $f_{lie}$ as in the mean teacher (MT) framework, we also apply EMA from $f_{lie}$ to $f_{sam}$, enabling collaborative learning and mutual teaching during training, resulting in a powerful encoder with superior feature representation. First, we train the medical branch with the supervised loss $\mathcal{L}_{med}$ (detail in the latter section). Then, the weights of $f_{lie}$ are updated with the EMA from $f_{sam}$ as 
\begin{equation}
   \theta_t^\prime = \alpha\theta_{t-1}^\prime + (1 - \alpha)\theta_t,
\end{equation}
where $\theta_{t-1}^\prime$ denotes the parameters of the network $f_{lie}$ at iteration $t - 1$, $\theta_t^\prime $ denotes the parameters of $f_lie$ after updating with EMA but before supervised training at iteration t, and $\alpha$ denotes the smoothing coefficient that is set to 0.99 \cite{he2023bilateral}.

Next, we train the general branch with supervised loss $\mathcal{L}_{gen}$ (detail in the latter section). Finally, we also update the $f_{sam}$ by the EMA of $f_{lie}$ as 
\begin{equation}
   \theta_t = \alpha\theta_{t-1} + (1 - \alpha)\theta_t^\prime,
\end{equation}
where $\theta_t$ denotes the parameters of the network $f_{sam}$ at iteration $t$, $\theta_{t}^\prime$ denotes the parameters of $f_{lie}$ at iteration t. It is worth noting that the respective branches in the first and second stages use the same network framework and the same supervised loss. With such a DSKD strategy, we can elegantly utilize the general knowledge of SAM for better feature representation.

\begin{figure}[!t]
\centering
\setlength{\belowcaptionskip}{-3mm}
\includegraphics[scale=0.45]{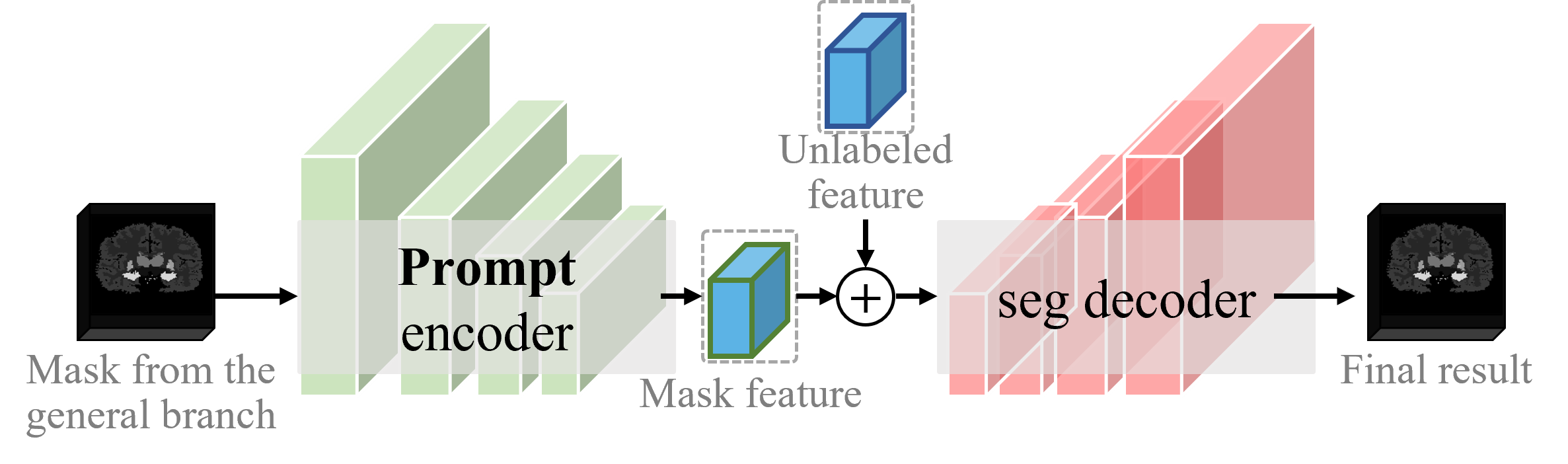}
\caption{The details of the proposed auto-prompting decoder}
\label{fig3}
\end{figure}

\subsection{Auto-Prompting Segmentation Decoder}
Current SAM-based medical architectures \cite{huang2024segment,ma2024segment,lin2023samus} and lightweight SAM \cite{zhang2023faster,zhou2023edgesam,xiong2024efficientsam} rely highly on user interactions, such as points, boxes, and masks, to help the model generate favorable results. However, user interactions need medical expertise, which is expensive and time-consuming, which is impractical for clinical scenarios. To alleviate the burden of the manual prompt, we propose the auto-prompting (AP) decoder to utilize the segmentation masks generated from the general branch for automatic purposes. As shown in Fig. \ref{fig3}, the general branch provides a coarse mask $\hat{y}_g$ that is fed into the prompt encoder to extract features. These features, along with the unlabeled image, are input into the segmentation decoder to generate the final predicted masks.

\subsection{Loss Function}



\subsubsection{Loss for registration.} To ensure smooth deformation, we apply a smoothness loss $\mathcal{L}_{smooth} = \sum{i\in\phi} ||\nabla\phi(i)||^2$, where $i$ represents the positions of the voxels in $\phi$ and $\nabla\phi(i)$ is its gradient. The similarity between the warped image and target image $y$ is measured by $\mathcal{L}_{sim} = l_{sim}(x_a \circ \phi, x_u)$, where $\circ$ denotes the warp operation. To further enhance the registration with anatomical structures, we incorporate segmentation results. Specifically, we use the Dice loss $\mathcal{L}_{dice} = l_{dice}(y_a \circ \phi, \hat{y}_u^m)$, comparing the warped segmentation map $y_a$ with the predicted segmentation $\hat{y}_u^t$.

The total registration loss is then a weighted sum of these components: \begin{equation} \mathcal{L}_{reg} = \lambda_1 \mathcal{L}_{smooth} + \lambda_2 \mathcal{L}_{sim} + \lambda_3 \mathcal{L}_{dice}. \end{equation}

\subsubsection{Loss for segmentation.} 
We adopt the same loss to train the segmentation task for both general and medical branches, denoting as $\mathcal{L}_{seg}^g$ and $\mathcal{L}_{seg}^m$, respectively. Specifically, the pseudo labels $y^i_p = y_a \circ \phi$ are generated by the registration task to supervise the segmentation task as 
\begin{equation}
    \mathcal{L}_{seg}^g = l_{dice}(y_a\circ\phi, \hat{y}_u^g)
\end{equation}
\begin{equation}
    \mathcal{L}_{seg}^m = l_{dice}(y_a\circ\phi, \hat{y}_u^m)
\end{equation}

\section{Experiments}

\subsection{Datasets}

We evaluated our method using two independent datasets: OASIS \cite{marcus2010open} and Lung-CT \cite{mackay2019self}, which were unseen during the pre-training phase. Data preprocessing is performed using FreeSurfer and FSL, including motion correction, NU intensity normalization, normalization, and affine registration. All scans are cropped and resampled to 128 × 128 × 128 with isotropic 1mm resolution. 


\textbf{\emph{OASIS. }}
The OASIS dataset focuses particularly on the early diagnosis of Alzheimer's Disease (AD), 
including healthy individuals and those with varying degrees of AD. 
For our research, the OASIS dataset is divided into 86 portions for training and 25 portions for testing.

\textbf{\emph{Lung-CT. }}
We downloaded CT lung images from Learn2reg Challenge, including 200 training images and 21 test images in size of 512 × 512 × 16 with segmentation annotations. We take expiration data as target images and inspiration as source images.



\subsection{Experimental Setup and Evaluation Metrics}


\subsubsection{Experimental Setup}
Our training process consists of pre-training via knowledge distillation with SAM to train a lightweight encoder and fine-tuning using a mutual EMA strategy. The framework is implemented using PyTorch on an NVIDIA A40 with 48 GB of RAM. Data augmentations include mirroring transformations, spatial transformations, rotation, and scaling. Both the registration and segmentation networks are trained using the Adam optimizer with a learning rate of 0.001 and batch size of 1. Additionally, we apply an EMA update strategy with a decay rate of 0.999 to smooth model weight updates.

\subsubsection{Evaluation Metrics}

To compare the performance of different methods and evaluate the performance of our proposed method, we use two common metrics: Dice and NCC. The Dice coefficient measures overlap between regions, with higher values indicating better segmentation. NCC quantifies the similarity between the deformed and target images, with higher values reflecting better alignment. A lower variance reveals our method is more stable and robust. 


\begin{table}[!t]
\caption{Quantitative comparison with various methods for segmentation task on the OASIS and CT-Lung datasets. The best performance is marked in bold.}
\renewcommand{\arraystretch}{1.2}
\begin{tabular}{c|>{\centering\arraybackslash}p{2.5cm}>{\centering\arraybackslash}p{2cm}|>{\centering\arraybackslash}p{2.5cm}>{\centering\arraybackslash}p{2cm}|cc}
\hline
\multirow{2}{*}{Methods} & \multicolumn{2}{c|}{MR-OASIS}                                  & \multicolumn{2}{c|}{CT-lung}                                   & \multirow{2}{*}{Prompt}& \multirow{2}{*}{Param(M)}\\ \cline{2-5}
                         & \multicolumn{1}{c|}{Dice(\%)}          & Ncc                   & \multicolumn{1}{c|}{Dice(\%)}          & Ncc                   &                             &                           \\ \hline
SST                      & \multicolumn{1}{c|}{76.5±0.7}          & \textbf{0.364±0.005}  & \multicolumn{1}{c|}{61.8±0.8}          & 0.352±0.020           & ×& 22.58                     \\
TBIOneShot               & \multicolumn{1}{c|}{80.8±0.9}          & 0.357±0.006           & \multicolumn{1}{c|}{74.5±1.3}          & 0.405±0.018           & × & 50.7                      \\
BRBS                     & \multicolumn{1}{c|}{80.3±1.1}          & 0.324±0.008           & \multicolumn{1}{c|}{89.4±0.4}          & 0.433±0.031           & ×& 5.89                      \\
Bi-JROS                  & \multicolumn{1}{c|}{81.4±0.9}          & 0.362±0.004           & \multicolumn{1}{c|}{90.5±0.9}          & 0.455±0.019           & ×& 5.02                      \\ \hline
SAM                      & \multicolumn{1}{c|}{32.4±1.0}                  & \multicolumn{1}{c|}{-} & \multicolumn{1}{c|}{89.5±1.5}                  & \multicolumn{1}{c|}{-} & \multicolumn{1}{c}{$\checkmark$}        & 357.64     \\
Med-SAM                  & \multicolumn{1}{c|}{80.1±1.1}          & -          & \multicolumn{1}{c|}{90.2±1.3}          & -          &                             $\checkmark$&  357.65                         \\   \hline
Ours                     & \multicolumn{1}{c|}{\textbf{82.4±0.9}} & 0.363±0.004           & \multicolumn{1}{c|}{\textbf{93.6±0.8}} & \textbf{0.458±0.012}  &× & 5.02                    \\ \hline
\end{tabular}
\label{tab1}
\end{table}

\begin{table}[!t]
\caption{Ablation study on dual-stage knowledge distillation and auto-prompting.}
\centering
\setlength{\tabcolsep}{1mm}{
\begin{tabular}{cccc|cc}
\toprule[0.3mm]
\multicolumn{1}{c}{\multirow{2}{*}{Baseline}} & \multicolumn{1}{c}{\multirow{2}{*}{\begin{tabular}[c]{@{}c@{}}One-\\ stage\end{tabular}}} & \multicolumn{1}{c}{\multirow{2}{*}{\begin{tabular}[c]{@{}c@{}}Dual-\\ stage\end{tabular}}} & \multirow{2}{*}{\begin{tabular}[c]{@{}l@{}}Prompt-\\ guided\end{tabular}} & \multirow{2}{*}{S-Dice(\%)}& \multirow{2}{*}{R-Dice(\%)} \\
\multicolumn{1}{c}{}                          & \multicolumn{1}{c}{}                                                                      & \multicolumn{1}{c}{}                                                                       &                                                                           &                               \\ \midrule
\multicolumn{1}{c}{$\surd$}                          & \multicolumn{1}{c}{}                                                                      & \multicolumn{1}{c}{}                                                                      & & 79.8 ± 0.7    & 78.5 ± 0.6                              \\ $\surd$
& {S→T}                                                                                      &                                                                                          & $\surd$ & 81.3 ± 1.1    & 78.5 ± 0.7  \\ 
$\surd$
& {T→S}                                                                                      &                                                                                         &$\surd$ & 81.9 ± 1.0    & 79.5 ± 0.9     \\$\surd$
& $\surd$                                                                                       & $\surd$                                                                                  &  & 80.8 ± 0.1                                                                          & 78.5 ± 0.1                              \\$\surd$
& $\surd$                                                                                         & $\surd$                                                                                          & $\surd$         &82.2 ± 0.9    & 80.1 ± 0.8           \\
\bottomrule[0.3mm]
\end{tabular}}
\label{tab3}
\end{table}

\subsection{Comparison Experiments}

\subsubsection{Quantitative comparison.} To evaluate the superiority of our method in the joint tasks of registration and segmentation, we compared it with several state-of-the-art methods, including SST \cite{tomar2022self}, TBIOneShot\cite{zhao2023one}, BRBS \cite{he2022learning}, Bi-JROS \cite{fan2024bi}, and SAM-based approaches (SAM \cite{kirillov2023segment} and Med-SAM \cite{ma2024segment}). For fairness, each method was trained from scratch using the same experimental setup and evaluated on unseen test sets, relying on only an atlas image. The SAM experiment used manual prompts consisting of a positive point, a negative point, and a bounding box. These prompts are derived in advance from the ground-truth masks. As shown in Table \ref{tab1}, with limited training and testing data, our method achieved the highest segmentation Dice scores of 82.4\% (NCC = 0.363) on the OASIS dataset and 93.6\% (NCC = 0.458) on the Lung-CT dataset. This shows that the proposed RRL-MedSAM outperforms other state-of-the-art methods and SAM-based approaches by effectively using SAM, providing more reliable segmentation results. Furthermore, our method offers advantages in speed and efficiency, requiring shorter training times and achieving rapid convergence.

\subsubsection{Qualitative comparison. }
\begin{figure}[!ht]
\centering
\setlength{\belowcaptionskip}{-3mm}
\includegraphics[scale=0.3]{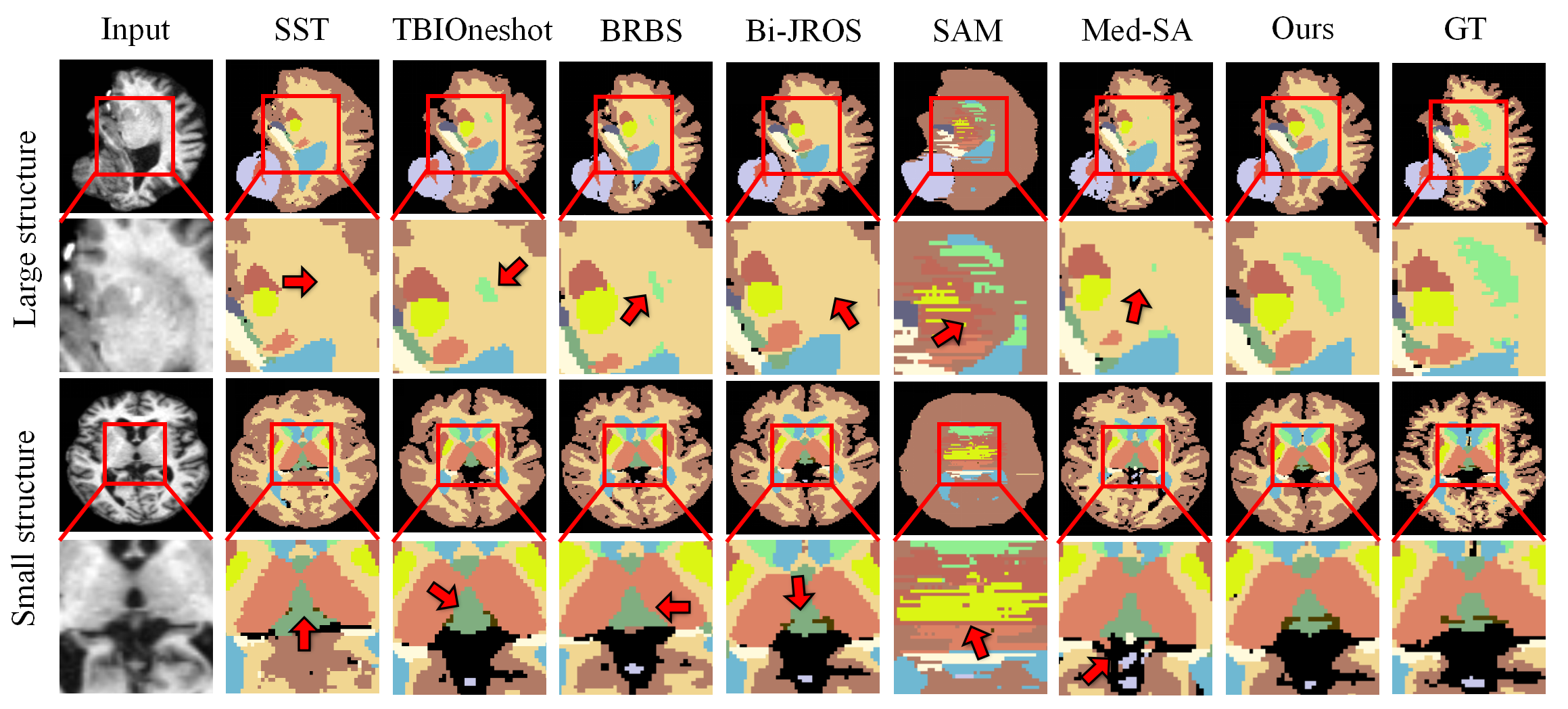}
\caption{Visualization of Brain Segmentation Results for Two Views： The segmentation results of larger brain structures and smaller cerebellar tissue 3rd/4th Ventricle (Ven). The red arrows point to the segmentation errors.}
\label{fig5}
\end{figure}

Fig. \ref{fig5} presents the segmentation and registration results of various JRS methods on the OASIS test dataset. In the segmentation task, our method demonstrated the fewest errors and the highest overlap with ground truth (GT). Compared to other methods, our algorithm performs better in preserving the integrity of tissue structures and clarity of boundaries. This is attributed to the optimized design in feature extraction and model training, which effectively captures key information in the images and reduces the interference of noise and artefacts during the segmentation process.

\subsubsection{Effectiveness of dual-stage knowledge distillation.} 
Table \ref{tab3} shows the results of ablation experiments on dual-stage knowledge distillation (DSKD) to assess its impact on model performance. Starting with a baseline model without EMA updates, we test two unidirectional EMA strategies: S→T (student to teacher) and T→S (teacher to student). The results indicate that both unidirectional updates improve registration and segmentation performance. Notably, the two-stage DSKD method achieves the best performance, demonstrating that our DSKD strategy can significantly enhance model performance, particularly in the segmentation task.

\subsubsection{Effectiveness of Auto-prompting segmentation decoder.} To validate the effectiveness of the proposed AP decoder, we train the model without the prompt-guided strategy. As shown in Table \ref{tab3}, the without prompt-guided model achieves the Dice score of 80.8\% and 79.5\% in two tasks, which is lower 1.4\% and 1.6\% than our final model, respectively. This verifies the effectiveness of the prompt-guided strategy in the AP segmentation decoder, which will boost both the segmentation and registration performance. 


\section{Conclusion}
In this paper, we present a novel SAM-guided robust representation learning framework, named RRL-MedSAM, to adapt SAM to one-shot 3D medical image segmentation. Our approach leverages SAM's strong generalization ability through a dual-stage knowledge distillation (DSKD) strategy, which first distills general knowledge from natural to medical images to train a lightweight encoder, and then applies mutual exponential moving average (mutual-EMA) updates for the general and medical-specific encoders. Additionally, we introduce an auto-prompting (AP) segmentation decoder to adopt the mask generated from the lightweight model as an auto-prompt to enhance segmentation performance. Extensive experiments demonstrate that RRL-MedSAM achieves state-of-the-art performance on two public benchmarks for segmentation tasks.


\bibliographystyle{splncs04}
\bibliography{ref}

\begin{thebibliography}{10}
\providecommand{\url}[1]{\texttt{#1}}
\providecommand{\urlprefix}{URL }
\providecommand{\doi}[1]{https://doi.org/#1}

\bibitem{andresen2022deep}
Andresen, J., Kepp, T., Ehrhardt, J., Burchard, C.v.d., Roider, J., Handels,
  H.: Deep learning-based simultaneous registration and unsupervised
  non-correspondence segmentation of medical images with pathologies.
  International Journal of Computer Assisted Radiology and Surgery
  \textbf{17}(4),  699--710 (2022)

\bibitem{ding2021modeling}
Ding, Y., Yu, X., Yang, Y.: Modeling the probabilistic distribution of
  unlabeled data for one-shot medical image segmentation. In: Proceedings of
  the AAAI conference on artificial intelligence. vol.~35, pp. 1246--1254
  (2021)

\bibitem{elmahdy2021joint}
Elmahdy, M.S., Beljaards, L., Yousefi, S., Sokooti, H., Verbeek, F., Van
  Der~Heide, U.A., Staring, M.: Joint registration and segmentation via
  multi-task learning for adaptive radiotherapy of prostate cancer. IEEE Access
   \textbf{9},  95551--95568 (2021)

\bibitem{fan2024bi}
Fan, X., Wang, X., Gao, J., Wang, J., Luo, Z., Liu, R.: Bi-level learning of
  task-specific decoders for joint registration and one-shot medical image
  segmentation. In: Proceedings of the IEEE/CVF Conference on Computer Vision
  and Pattern Recognition. pp. 11726--11735 (2024)

\bibitem{gong20233dsam}
Gong, S., Zhong, Y., Ma, W., Li, J., Wang, Z., Zhang, J., Heng, P.A., Dou, Q.:
  3dsam-adapter: Holistic adaptation of sam from 2d to 3d for promptable
  medical image segmentation. arXiv preprint arXiv:2306.13465  (2023)

\bibitem{he2023bilateral}
He, A., Li, T., Yan, J., Wang, K., Fu, H.: Bilateral supervision network for
  semi-supervised medical image segmentation. IEEE Transactions on Medical
  Imaging  (2023)

\bibitem{he2022learning}
He, Y., Ge, R., Qi, X., Chen, Y., Wu, J., Coatrieux, J.L., Yang, G., Li, S.:
  Learning better registration to learn better few-shot medical image
  segmentation: Authenticity, diversity, and robustness. IEEE Transactions on
  Neural Networks and Learning Systems  \textbf{35}(2),  2588--2601 (2022)

\bibitem{huang2024segment}
Huang, Y., Yang, X., Liu, L., Zhou, H., Chang, A., Zhou, X., Chen, R., Yu, J.,
  Chen, J., Chen, C., et~al.: Segment anything model for medical images?
  Medical Image Analysis  \textbf{92},  103061 (2024)

\bibitem{kirillov2023segment}
Kirillov, A., Mintun, E., Ravi, N., Mao, H., Rolland, C., Gustafson, L., Xiao,
  T., Whitehead, S., Berg, A.C., Lo, W.Y., et~al.: Segment anything. In:
  Proceedings of the IEEE/CVF International Conference on Computer Vision. pp.
  4015--4026 (2023)

\bibitem{lin2023samus}
Lin, X., Xiang, Y., Zhang, L., Yang, X., Yan, Z., Yu, L.: Samus: Adapting
  segment anything model for clinically-friendly and generalizable ultrasound
  image segmentation. arXiv preprint arXiv:2309.06824  (2023)

\bibitem{ma2024segment}
Ma, J., He, Y., Li, F., Han, L., You, C., Wang, B.: Segment anything in medical
  images. Nature Communications  \textbf{15}(1), ~654 (2024)

\bibitem{mackay2019self}
MacKay, M., Vicol, P., Lorraine, J., Duvenaud, D., Grosse, R.: Self-tuning
  networks: Bilevel optimization of hyperparameters using structured
  best-response functions. arXiv preprint arXiv:1903.03088  (2019)

\bibitem{marcus2010open}
Marcus, D.S., Fotenos, A.F., Csernansky, J.G., Morris, J.C., Buckner, R.L.:
  Open access series of imaging studies: longitudinal mri data in nondemented
  and demented older adults. Journal of cognitive neuroscience
  \textbf{22}(12),  2677--2684 (2010)

\bibitem{sinclair2022atlas}
Sinclair, M., Schuh, A., Hahn, K., Petersen, K., Bai, Y., Batten, J., Schaap,
  M., Glocker, B.: Atlas-istn: joint segmentation, registration and atlas
  construction with image-and-spatial transformer networks. Medical Image
  Analysis  \textbf{78},  102383 (2022)

\bibitem{sun2023dual}
Sun, Y., Wang, F., Shu, J., Wang, H., Wang, L., Meng, D., Lian, C.: Dual
  meta-learning with longitudinally generalized regularization for one-shot
  brain tissue segmentation across the human lifespan. In: 2023 IEEE/CVF
  International Conference on Computer Vision (ICCV). pp. 21061--21071. IEEE
  (2023)

\bibitem{tomar2022self}
Tomar, D., Bozorgtabar, B., Lortkipanidze, M., Vray, G., Rad, M.S., Thiran,
  J.P.: Self-supervised generative style transfer for one-shot medical image
  segmentation. In: Proceedings of the IEEE/CVF winter conference on
  applications of computer vision. pp. 1998--2008 (2022)

\bibitem{vandewinckele2020segmentation}
Vandewinckele, L., Willems, S., Robben, D., Van Der~Veen, J., Crijns, W.,
  Nuyts, S., Maes, F.: Segmentation of head-and-neck organs-at-risk in
  longitudinal ct scans combining deformable registrations and convolutional
  neural networks. Computer Methods in Biomechanics and Biomedical Engineering:
  Imaging \& Visualization  \textbf{8}(5),  519--528 (2020)

\bibitem{wu2023medical}
Wu, J., Ji, W., Liu, Y., Fu, H., Xu, M., Xu, Y., Jin, Y.: Medical sam adapter:
  Adapting segment anything model for medical image segmentation. arXiv
  preprint arXiv:2304.12620  (2023)

\bibitem{xiong2024efficientsam}
Xiong, Y., Varadarajan, B., Wu, L., Xiang, X., Xiao, F., Zhu, C., Dai, X.,
  Wang, D., Sun, F., Iandola, F., et~al.: Efficientsam: Leveraged masked image
  pretraining for efficient segment anything. In: Proceedings of the IEEE/CVF
  Conference on Computer Vision and Pattern Recognition. pp. 16111--16121
  (2024)

\bibitem{xu2024esp}
Xu, Q., Li, J., He, X., Liu, Z., Chen, Z., Duan, W., Li, C., He, M.M., Tesema,
  F.B., Cheah, W.P., et~al.: Esp-medsam: Efficient self-prompting sam for
  universal domain-generalized medical image segmentation. arXiv preprint
  arXiv:2407.14153  (2024)

\bibitem{zhang2023faster}
Zhang, C., Han, D., Qiao, Y., Kim, J.U., Bae, S.H., Lee, S., Hong, C.S.: Faster
  segment anything: Towards lightweight sam for mobile applications. arXiv
  preprint arXiv:2306.14289  (2023)

\bibitem{zhang2024efficientvit}
Zhang, Z., Cai, H., Han, S.: Efficientvit-sam: Accelerated segment anything
  model without performance loss. arXiv preprint arXiv:2402.05008  (2024)

\bibitem{zhao2021deep}
Zhao, F., Wu, Z., Wang, L., Lin, W., Xia, S., Li, G., Consortium, U.B.C.P.: A
  deep network for joint registration and parcellation of cortical surfaces.
  In: Medical Image Computing and Computer Assisted Intervention--MICCAI 2021:
  24th International Conference, Strasbourg, France, September 27--October 1,
  2021, Proceedings, Part IV 24. pp. 171--181. Springer (2021)

\bibitem{zhao2023one}
Zhao, X., Shen, Z., Chen, D., Wang, S., Zhuang, Z., Wang, Q., Zhang, L.:
  One-shot traumatic brain segmentation with adversarial training and
  uncertainty rectification. In: International Conference on Medical Image
  Computing and Computer-Assisted Intervention. pp. 120--129. Springer (2023)

\bibitem{zhou2023edgesam}
Zhou, C., Li, X., Loy, C.C., Dai, B.: Edgesam: Prompt-in-the-loop distillation
  for on-device deployment of sam. arXiv preprint arXiv:2312.06660  (2023)

\end{thebibliography}

\end{document}